\documentstyle[epsfig,a4p]{article}

\begin{document}


\newcommand{\VMtau}     {\mbox{$1777.0 \pm 0.3$~MeV/c$^2$}}
\newcommand{\VMb}       {\mbox{$4.13 \pm 0.06$~GeV}}
\newcommand{\VMc}     {\mbox{$1.31 \pm 0.06$~GeV}}
\newcommand{\VMz}       {\mbox{$91.187 \pm 0.007$~GeV}}

\newcommand{\Vnelec}      {\mbox{$33\;073$}}
\newcommand{\Vntau}      {\mbox{$186\;197$}}

\newcommand{\Vpstat}      {\mbox{$0.089$}}
\newcommand{\Vpsys}      {\mbox{$0.057$}}
\newcommand{\Vpbr}      {\mbox{$17.806$}}

\newcommand{\Vstat}      {\mbox{$0.09$}}
\newcommand{\Vsys}      {\mbox{$0.06$}}
\newcommand{\Vbr}      {\mbox{$17.81$}}

\newcommand{\VMCmod}       {\mbox{$0.031$}}
\newcommand{\VPresys}      {\mbox{$0.015$}}
\newcommand{\VPhosys}      {\mbox{$0.010$}}
\newcommand{\VMCstat}      {\mbox{$0.015$}}
\newcommand{\Vntbgsys}     {\mbox{$0.025$}}
\newcommand{\Vbiassys}     {\mbox{$0.034$}}
\newcommand{\Vbias}      {\mbox{$1.0009 \pm0.0019$}}

\newcommand{\Veff}      {\mbox{$98.40\pm 0.05$}}
\newcommand{\Vtback}      {\mbox{$1.21\pm0.04$}}
\newcommand{\Vbackone}      {\mbox{$0.48\pm0.03$}}
\newcommand{\Vbacktwo}      {\mbox{$0.69\pm0.03$}}
\newcommand{\Vbackthree}      {\mbox{$0.04\pm0.01$}}

\newcommand{\Vbhabha}      {\mbox{$0.15 \pm 0.03$}}
\newcommand{\Vmumu}      {\mbox{$0.52\pm0.03$}}
\newcommand{\Vqqbar}      {\mbox{$0.38 \pm 0.05$}}
\newcommand{\Veemm}      {\mbox{$0.06 \pm 0.02$}}
\newcommand{\Veeee}      {\mbox{$0.11 \pm 0.01$}}
\newcommand{\Veett}      {\mbox{$0.045 \pm 0.01$}}
\newcommand{\Veell}      {\mbox{$0.21 \pm 0.02$}}
\newcommand{\Vntback}      {\mbox{$1.26 \pm 0.07$}}

\newcommand{\Vebhabha}      {\mbox{$0.67 \pm 0.10$}}
\newcommand{\Veeeee}      {\mbox{$0.59 \pm 0.07$}}
\newcommand{\Veeell}      {\mbox{$0.62 \pm 0.07$}}
\newcommand{\Vnteback}      {\mbox{$1.29 \pm 0.12$}}

\newcommand{\VRtau}       {\mbox{$3.642$}}
\newcommand{\VRterr}     {\mbox{$0.033$}}

\newcommand{\Vastau}      {\mbox{$0.334$}}
\newcommand{\Vasteexp}      {\mbox{$0.010$}}

\newcommand{\Vasz}        {\mbox{$0.1204$}}
\newcommand{\Vaszeexp}     {\mbox{$0.0011$}}

\newcommand{\Vnpertcor}   {\mbox{$-0.020\pm 0.004$}}
\newcommand{\Vnpert}   {\mbox{$-0.010\pm 0.004$}}
\newcommand{\VDtwocor}      {\mbox{$-0.010 \pm 0.002$}}
\newcommand{\VDfourcor}      {\mbox{$-0.0033 \pm 0.0005$}}
\newcommand{\VDsixcor}      {\mbox{$-0.007 \pm 0.004$}}
\newcommand{\VDeightcor}      {\mbox{$\approx 1 \times 10^{-5}$}}

\newcommand{\Vvud}      {\mbox{$0.9753 \pm 0.0008$}}
\newcommand{\Vvus}      {\mbox{$0.2205 \pm 0.0018$}}
\newcommand{\Vsew}      {\mbox{$1.0194 \pm 0.0040$}}
\newcommand{\Vdew}      {\mbox{$0.0010$}}

\newcommand{\Vtew}      {\mbox{$0.004$}}
\newcommand{\Vzew}      {\mbox{$0.0005$}}

\newcommand{\Vtsd}      {\mbox{$0.009$}}
\newcommand{\Vzsd}      {\mbox{$0.0009$}}

\newcommand{\Vtrs}      {\mbox{$0.010$}}
\newcommand{\Vzrs}      {\mbox{$0.0011$}}

\newcommand{\Vtnp}      {\mbox{$0.004$}}
\newcommand{\Vznp}      {\mbox{$0.0005$}}

\newcommand{\Vtkf}      {\mbox{$0.006$}}
\newcommand{\Vzkf}      {\mbox{$0.0007$}}

\newcommand{\Vzrun}      {\mbox{$0.0003$}}

\newcommand{\Vasteth}      {\mbox{$0.016$}}
\newcommand{\Vaszeth}      {\mbox{$0.0019$}}


\newcommand{\VunivRC}      {\mbox{$0.9960$}}
\newcommand{\VObrtmu}      {\mbox{$(17.36 \pm 0.27)\%$}}
\newcommand{\VOPNbrtmu}      {\mbox{$(17.48 \pm 0.12 \pm 0.08)\%$}}
\newcommand{\VOtlife}      {\mbox{$289.2 \pm 1.7 \pm 1.2$ fs}}
\newcommand{\Vmulife}      {\mbox{$2.19703 \pm 0.00004$ $\mu$s}}

\newcommand{\Vgmuge}      {\mbox{$1.0011 \pm 0.0083$}}
\newcommand{\VPNgmuge}      {\mbox{$1.0046 \pm 0.0051$}}
\newcommand{\Vgtaugmu}      {\mbox{$1.0025 \pm 0.0047$}}

\newcommand{\beq}       {\begin{equation}}
\newcommand{\eeq}       {\end{equation}}
\newcommand{\etal}      {{\it et~al.}}
\newcommand{\mperiod}      {\mbox{\hspace{0.4cm}.}}
\newcommand{\mcomma}      {\mbox{\hspace{0.4cm},}}

\newcommand{\as}       {\mbox{$\alpha_s$}}
\newcommand{\asmt}       {\mbox{$\alpha_s(m_\tau^2)$}}
\newcommand{\asmz}       {\mbox{$\alpha_s(m_{\rm Z}^2)$}}
\newcommand{\nf}       {\mbox{$n_f$}}
\newcommand{\rtau}       {\mbox{$R_\tau$}}
\newcommand{\vud}       {\mbox{$V_{\mathrm{ud}}$}}
\newcommand{\vus}       {\mbox{$V_{\mathrm{us}}$}}

\newcommand{\piz}{\mbox{$\pi^0$}}
\newcommand{\pic}{\mbox{$\pi^{\pm}$}}
\newcommand{\hpiz}{\mbox{$h\piz$}}

\newcommand{\Wpm}        {\mbox{$\mathrm{W}^{\pm}$}}
\newcommand{\zo}        {\mbox{$\mathrm{Z}^0$}}
\newcommand{\Ztt}	{\mbox{$\zo \rightarrow \tau^+ \tau^-$}}
\newcommand{\tautau}	{\mbox{$\tau^+ \tau^-$}}
\newcommand{\mumu}	{\mbox{$\mu^+ \mu^-$}}
\newcommand{\qqbar}	{\mbox{$\mathrm{q}\bar{\mathrm{q}}$}}
\newcommand{\epem}	{\mbox{$\mathrm{e}^+\mathrm{e}^-$}}

\newcommand{\thmiss}    {\mbox{$\theta_{miss}$}}
\newcommand{\cosmiss}   {\mbox{$| \cos \thmiss |$}}
\newcommand{\ecm}      {\mbox{$E_{\mathrm{cm}}$}}
\newcommand{\ebeam}      {\mbox{$E_{\mathrm{beam}}$}}
\newcommand{\evis}      {\mbox{$E_{\mathrm{vis}}$}}
\newcommand{\etot}      {\mbox{$E_{\mathrm{tot}}$}}
\newcommand{\ptot}      {\mbox{$p_{\mathrm{tot}}$}}
\newcommand{\ptrk}      {\mbox{$p$}}
\newcommand{\ecls}      {\mbox{$E$}}
\newcommand{\ncls}      {\mbox{$N_{\mathrm{clus}}$}}
\newcommand{\costav}   {\mbox{$\overline{| \cos \theta |}$}}

\newcommand{\dedx}      {{\mathrm{d}} E/{\mathrm{d}}x}

\newcommand{\denor}      {\mbox{$N(\dedx)$}}
\newcommand{\norep}      {\mbox{$N(E/p)$}}
\newcommand{\dtheta}     {\mbox{$N(\Delta\theta)$}}
\newcommand{\dphi}       {\mbox{$N(\Delta\phi)$}}
\newcommand{\eneut}      {\mbox{$N_{\mathrm{neut}}$}}
\newcommand{\hcal}      {\mbox{$N_{\mathrm{HCAL}}$}}
\newcommand{\ntrks}      {\mbox{$N_{\mathrm{tracks}}$}}
\newcommand{\nacl}      {\mbox{$N_{\mathrm{ACl}}$}}

\newcommand{\phisect}   {\mbox{$\phi_{\mathrm{sector}}$}}
\newcommand{\acolin}    {\mbox{$\theta_{\mathrm{acol}}$}}

\newcommand{\mue}      {\mbox{$\mu^{-}\rightarrow\mathrm{e}^{-}
                                          \bar{\nu}_{\mathrm{e}} \nu_\mu$}}

\newcommand{\tl}      {\mbox{$\tau^{-}\rightarrow l^{-}
                                          \bar{\nu}_l \nu_\tau$}}

\newcommand{\te}      {\mbox{$\tau^{-}\rightarrow \mathrm{e}^{-}
                                          \bar{\nu}_{\mathrm{e}} \nu_\tau$}}
\newcommand{\tmu}     {\mbox{$\tau^{-}\rightarrow\mu^{-}
                                          \bar{\nu}_{\mu}\nu_\tau$}}

\newcommand{\tpi}     {\mbox{$\tau^{-}\rightarrow \pi^{-}\nu_\tau$}}
\newcommand{\tpigopiz}  {\mbox{$\tau^{-}\rightarrow\pi^{-}(\geq1 \pi^0)\nu_\tau$}}
\newcommand{\threepi}     {\mbox{$\tau^{-}\rightarrow  2\pi^{-} \pi^{+}\nu_\tau$}}

\newcommand{\tpinpiz}  {\mbox{$\tau^{-}\rightarrow\pi^{-}(\geq0 \pi^0)\nu_\tau$}}

\newcommand{\thad}     {\mbox{$\tau^{-}\rightarrow \mathrm{ hadrons~} \nu_\tau$}}

\newcommand{\Necand}    {\mbox{$N_{\mathrm{e}}$}}
\newcommand{\Ntcand}    {\mbox{$N_{\tau}$}}
\newcommand{\eff}       {\mbox{$\epsilon_{\mathrm{e}}$}}
\newcommand{\fe}        {\mbox{$f_{\tau\rightarrow \mathrm{e}\bar{\nu}\nu}$}}
\newcommand{\fet}        {\mbox{$f_{\tau\rightarrow\mathrm{e}\bar{\nu}\nu}^\tau$}}
\newcommand{\fent}        {\mbox{$f_{\tau\rightarrow\mathrm{e}\bar{\nu}\nu}^{\mathrm{non-}\tau}$}}
\newcommand{\ftau}      {\mbox{$f_\tau$}}
\newcommand{\bias}      {\mbox{$F_{\mathrm{B}}$}}
\newcommand{\brte}      {\mbox{B($\tau^{-}\rightarrow \mathrm{e}^{-}
                                          \bar{\nu}_{\mathrm{e}} \nu_\tau$)}}
\newcommand{\brmue}      {\mbox{B($\mu^{-}\rightarrow \mathrm{e}^{-}
                                          \bar{\nu}_{\mathrm{e}} \nu_\mu$)}}
\newcommand{\brtmu}      {\mbox{B($\tau^{-}\rightarrow \mu^{-}
                                          \bar{\nu}_\mu \nu_\tau$)}}
\newcommand{\ee}  {\mbox{$\mathrm{e}^+\mathrm{e}^-$}}
\newcommand{\qq}  {\mbox{$\mathrm{q}^+\bar{\mathrm{q}}$}}

\newcommand{\eell} {\mbox{$\mathrm{e}^+\mathrm{e}^- \rightarrow \mathrm{l}^+ \mathrm{l}^-$ }}
\newcommand{\eeee} {\mbox{$\mathrm{e}^+\mathrm{e}^- \rightarrow \mathrm{e}^+ \mathrm{e}^-$}}
\newcommand{\eemm} {\mbox{$\mathrm{e}^+\mathrm{e}^- \rightarrow \mu^+ \mu^-$}}
\newcommand{\eett} {\mbox{$\mathrm{e}^+\mathrm{e}^- \rightarrow \tau^+ \tau^-$}}
\newcommand{\eeqq} {\mbox{$\mathrm{e}^+\mathrm{e}^- \rightarrow \mathrm{q}\bar{\mathrm{q}}$}}
\newcommand{\eeeell} {\mbox{$\mathrm{e}^+\mathrm{e}^- \rightarrow (\mathrm{e}^+ \mathrm{e}^-)\mathrm{l}^+\mathrm{l}^-$}}
\newcommand{\eeeeee} {\mbox{$\mathrm{e}^+\mathrm{e}^- \rightarrow (\mathrm{e}^+ \mathrm{e}^-)\mathrm{e}^+\mathrm{e}^-$}}
\newcommand{\eeeemm} {\mbox{$\mathrm{e}^+\mathrm{e}^- \rightarrow (\mathrm{e}^+ \mathrm{e}^-)\mu^+\mu^-$}}
\newcommand{\eeeett} {\mbox{$\mathrm{e}^+\mathrm{e}^- \rightarrow (\mathrm{e}^+ \mathrm{e}^-)\tau^+\tau^-$}}

\newcommand{\netot} {\mbox{$N_{\tau \rightarrow \mathrm{e} \nu \bar{\nu}}^{\mathrm{before}}$}}
\newcommand{\nttot} {\mbox{$N_{\tau}^{\mathrm{before}}$}}
\newcommand{\nesel} {\mbox{$N_{\tau \rightarrow \mathrm{e} \nu \bar{\nu}}^{\mathrm{after}}$}}
\newcommand{\ntsel} {\mbox{$N_{\tau}^{\mathrm{after}}$}}


\begin{titlepage}
\begin{center}{\large   EUROPEAN LABORATORY FOR PARTICLE PHYSICS
}\end{center}\bigskip
\begin{flushright}
CERN-EP/98-175 \\ 6th November 1998
\end{flushright}
\bigskip\bigskip\bigskip\bigskip\bigskip
\begin{center}{\huge\bf   \boldmath A Measurement of the  \\  \te\ Branching Ratio   \\
}\end{center}\bigskip\bigskip
\begin{center}{\LARGE The OPAL Collaboration
}\end{center}\bigskip\bigskip
\bigskip\begin{center}{\large  Abstract}\end{center}
The branching ratio for the decay \te\ has been measured using \zo\ decay data collected
by the OPAL experiment at LEP.  
  In total \Vnelec\ \te\ candidates were identified from a sample of \Vntau\ 
selected $\tau$ decays, giving a branching ratio of
 $\brte = (\Vbr \pm \Vstat \mbox{ (stat)} \pm  \Vsys \mbox{ (syst)} ) \%$.  
This result is combined with other
measurements to test e - $\mu$ and $\mu$ - $\tau$ universality in charged-current weak 
interactions.  Additionally, the strong coupling constant \asmt\ has been extracted
 from \brte\ and evolved to the \zo\ mass scale, giving 
$\asmz = \Vasz \pm \Vaszeexp \mbox{ (exp)} \pm \Vaszeth \mbox{ (theory)}$. 
\bigskip\bigskip\bigskip\bigskip
\bigskip\bigskip
\begin{center}{\large
(Submitted to Physics Letters B) \\
\bigskip\bigskip
}\end{center}
\end{titlepage}
\begin{center}{\Large        The OPAL Collaboration
}\end{center}\bigskip
\begin{center}{
G.\thinspace Abbiendi$^{  2}$,
K.\thinspace Ackerstaff$^{  8}$,
G.\thinspace Alexander$^{ 23}$,
J.\thinspace Allison$^{ 16}$,
N.\thinspace Altekamp$^{  5}$,
K.J.\thinspace Anderson$^{  9}$,
S.\thinspace Anderson$^{ 12}$,
S.\thinspace Arcelli$^{ 17}$,
S.\thinspace Asai$^{ 24}$,
S.F.\thinspace Ashby$^{  1}$
% ...
}\end{center}\bigskip
\bigskip
$^{  1}$School of Physics and Astronomy, University of Birmingham,
Birmingham B15 2TT, UK
\newline
$^{  2}$Dipartimento di Fisica dell' Universit\`a di Bologna and INFN,
I-40126 Bologna, Italy
\newline
$^{  3}$Physikalisches Institut, Universit\"at Bonn,
D-53115 Bonn, Germany
\newline
$^{  4}$Department of Physics, University of California,
Riverside CA 92521, USA
\newline
$^{  5}$Cavendish Laboratory, Cambridge CB3 0HE, UK
\newline
$^{  6}$Ottawa-Carleton Institute for Physics,
Department of Physics, Carleton University,
Ottawa, Ontario K1S 5B6, Canada
\newline
$^{  7}$Centre for Research in Particle Physics,
Carleton University, Ottawa, Ontario K1S 5B6, Canada
\newline
$^{  8}$CERN, European Organisation for Particle Physics,
CH-1211 Geneva 23, Switzerland
\newline
$^{  9}$Enrico Fermi Institute and Department of Physics,
University of Chicago, Chicago IL 60637, USA
\newline
$^{ 10}$Fakult\"at f\"ur Physik, Albert Ludwigs Universit\"at,
D-79104 Freiburg, Germany
\newline
$^{ 11}$Physikalisches Institut, Universit\"at
Heidelberg, D-69120 Heidelberg, Germany
\newline
$^{ 12}$Indiana University, Department of Physics,
Swain Hall West 117, Bloomington IN 47405, USA
\newline
$^{ 13}$Queen Mary and Westfield College, University of London,
London E1 4NS, UK
\newline
$^{ 14}$Technische Hochschule Aachen, III Physikalisches Institut,
Sommerfeldstrasse 26-28, D-52056 Aachen, Germany
\newline
$^{ 15}$University College London, London WC1E 6BT, UK
\newline
$^{ 16}$Department of Physics, Schuster Laboratory, The University,
Manchester M13 9PL, UK
\newline
$^{ 17}$Department of Physics, University of Maryland,
College Park, MD 20742, USA
\newline
$^{ 18}$Laboratoire de Physique Nucl\'eaire, Universit\'e de Montr\'eal,
Montr\'eal, Quebec H3C 3J7, Canada
\newline
$^{ 19}$University of Oregon, Department of Physics, Eugene
OR 97403, USA
\newline
$^{ 20}$CLRC Rutherford Appleton Laboratory, Chilton,
Didcot, Oxfordshire OX11 0QX, UK
\newline
$^{ 22}$Department of Physics, Technion-Israel Institute of
Technology, Haifa 32000, Israel
\newline
$^{ 23}$Department of Physics and Astronomy, Tel Aviv University,
Tel Aviv 69978, Israel
\newline
$^{ 24}$International Centre for Elementary Particle Physics and
Department of Physics, University of Tokyo, Tokyo 113-0033, and
Kobe University, Kobe 657-8501, Japan
\newline
$^{ 25}$Institute of Physical and Environmental Sciences,
Brunel University, Uxbridge, Middlesex UB8 3PH, UK
\newline
$^{ 26}$Particle Physics Department, Weizmann Institute of Science,
Rehovot 76100, Israel
\newline
$^{ 27}$Universit\"at Hamburg/DESY, II Institut f\"ur Experimental
Physik, Notkestrasse 85, D-22607 Hamburg, Germany
\newline
$^{ 28}$University of Victoria, Department of Physics, P O Box 3055,
Victoria BC V8W 3P6, Canada
\newline
$^{ 29}$University of British Columbia, Department of Physics,
Vancouver BC V6T 1Z1, Canada
\newline
$^{ 30}$University of Alberta,  Department of Physics,
Edmonton AB T6G 2J1, Canada
\newline
$^{ 31}$Duke University, Dept of Physics,
Durham, NC 27708-0305, USA
\newline
$^{ 32}$Research Institute for Particle and Nuclear Physics,
H-1525 Budapest, P O  Box 49, Hungary
\newline
$^{ 33}$Institute of Nuclear Research,
H-4001 Debrecen, P O  Box 51, Hungary
\newline
$^{ 34}$Ludwigs-Maximilians-Universit\"at M\"unchen,
Sektion Physik, Am Coulombwall 1, D-85748 Garching, Germany
\newline
\bigskip\newline
$^{  a}$ and at TRIUMF, Vancouver, Canada V6T 2A3
\newline
$^{  b}$ and Royal Society University Research Fellow
\newline
$^{  c}$ and Institute of Nuclear Research, Debrecen, Hungary
\newline
$^{  d}$ and Department of Experimental Physics, Lajos Kossuth
University, Debrecen, Hungary
\newline
$^{  e}$ on leave of absence from the University of Freiburg
\newline

\newpage
\section{Introduction \label{sec:intro}}

Decays of $\tau$ leptons probe the 
Standard Model in both the electroweak and strong sectors.  The leptonic branching ratios, 
in conjunction with $\tau$ lifetime and mass measurements, can be used to test the universality of 
charged-current couplings to electrons, muons and $\tau$ leptons~\cite{tau96_pich}. 
In addition, large QCD corrections to the $\tau$ decay width enable the strong coupling constant $\as$  to be determined at 
the $\tau$ mass scale from a measurement of the \te\ branching ratio~\cite{tau96_pich,tau96_braaten}.

This paper presents a measurement of the \te\ branching ratio\footnote{Charge conjugation is assumed
 throughout this paper.} using data collected by the OPAL experiment~\cite{opal_det}
 at energies near the \zo\ resonance.  
 This measurement uses a larger data set than earlier OPAL
measurements of $\brte$~\cite{randy,clayton}  and is based on a likelihood selection procedure.
Tests of e - $\mu$ and $\mu$ - $\tau$ universality are presented and a value of \asmt\  is
 extracted from the branching ratio result within the context of fixed-order perturbation theory
and evolved to the \zo\ mass.  

\section{Selection of {\boldmath $\tau$} jets \label{sec:tausel}}

The OPAL experiment collected data corresponding to approximately 170~pb$^{-1}$ 
of integrated luminosity during the period in which the LEP collider operated at 
centre-of-mass energies close to the \zo\ mass.
Approximately $90\%$ of these data were collected at the \zo\ peak and the remainder
at energies within about 2~GeV of the peak.  
Candidate $\tau$ events used in this analysis are selected from this data set
using a procedure similar to that described in previous OPAL publications~\cite{tp}.

Selection efficiencies and kinematic variable distributions were modelled
 using Monte Carlo simulated $\tautau$ event samples generated
with the KORALZ 4.02~\cite{koralz} package and the TAUOLA 2.0~\cite{tauola} library.  These
events were then passed through a full simulation of the OPAL detector~\cite{gopal}. 
Events generated at energies approximately 2~GeV above and below the \zo\ peak 
were combined with the events generated on-peak in proportion to the integrated luminosities of the data. 
Background contributions from non-$\tau$ sources were evaluated using Monte Carlo samples based on the following generators: 
Multihadronic events ($\eeqq$) were simulated using JETSET 7.4~\cite{jetset}, $\eemm$ 
events using KORALZ~\cite{koralz},  Bhabha events using BABAMC~\cite{babamc} and BHWIDE~\cite{bhwide},
and four fermion events  using VERMASEREN 1.01~\cite{verm}  and FERMISV~\cite{fermisv}.

At LEP, \zo\ bosons decaying at rest in the laboratory frame produce back-to-back
 $\tau$ pairs.   Each highly-relativistic $\tau$ subsequently decays in flight,  
 producing strongly collimated jets. 
 Events selected as $\tautau$ are required to have exactly two such jets
 identified using a cone algorithm~\cite{jetdef} with a cone half-angle of $35^\circ$.
  The average $|\cos \theta|$ of the two jets is required to be in the central region
 ($\overline{|\cos \theta |} <0.68$) of the OPAL detector\footnote{In the OPAL coordinate
 system the e$^-$ beam direction defines the $+z$ axis, and the centre of the LEP ring defines
 the $+x$ axis.  The polar angle $\theta$ is measured from the $+z$ axis, and the azimuthal 
angle $\phi$ is measured from the $+x$ axis}.  
The scalar sum of the momenta of all tracks, $\ptot$, and
the sum of the energies of all electromagnetic calorimeter (ECAL) clusters, 
$\etot$, are
required to satisfy $\evis \equiv \etot +\ptot >0.01 \cdot \ecm$, where \ecm\ is the
 centre-of-mass energy, in order to be considered a good event.  Backgrounds from cosmic rays are reduced to 
a negligible level by requirements on the time-of-flight detector.
Two-photon mediated four-fermion events, $\eeeell$ where 
$\mathrm{l}^+\mathrm{l}^-$ represents lepton pairs, are rejected by requirements on 
the acollinearity angle $\acolin$, defined as the supplement of the angle between
 the two jets, and on the total energy and momentum of tracks and clusters in the event.
  The remaining background from two-photon mediated events was evaluated
by comparing the \acolin\ distribution in data and simulation, and was estimated to be
    $(\Veell ) \%$. 

Decays of \zo\ bosons into lepton pairs produce low track and ECAL cluster multiplicities
 compared with multihadronic events ($\eeqq$).  The
 $\tautau$ candidates are required to possess between two and six tracks
 and no more than ten ECAL clusters. 
Jets produced in multihadronic events are typically less
 collimated than those resulting from $\tau$ decays. 
The residual \eeqq\ background was therefore evaluated by examining
 the collimation of tracks in the jets and was estimated to be $( \Vqqbar )\%$

Events within this low-multiplicity sample are expected to be lepton pairs, and are classified 
as either $\ee$, $\mu^+\mu^-$ or $\tau^+\tau^-$ based on tracking, calorimetry and muon chamber information. 
Bhabha events ($\eeee$) typically possess two high-momentum tracks and ECAL energy 
 close to the full centre-of-mass energy.
 These events are rejected by requiring  $\etot+0.3 \cdot \ptot \leq \ecm $ and $\evis \leq 1.4 \cdot \ecm$.
The residual Bhabha background in the $\tau$ sample was determined to be $(\Vbhabha)\%$ by examining 
the distribution of $\etot+0.3 \cdot \ptot$ in events containing an electron.  

Muon-pair events ($\eemm$) are characterized by two back-to-back
 high-momentum tracks associated with activity in the muon chambers or the hadron
 calorimeter and with little energy  in the
 ECAL.  Events with $\evis>0.6\cdot\ecm$ in combination with a
 muon tag~\cite{tp} in both jets are rejected.   Muon-pair events which survive this cut generally
do so because one of the muons has passed through a region of the detector which lacks muon chamber coverage.
The background due to muon pairs was evaluated by examining the \evis\ distribution of events with 
a single track in each jet, and was estimated to be $(\Vmumu )\%$.

 Fiducial cuts are imposed on the individual $\tau$ candidate jets to
 avoid regions of the detector associated with small gaps between structural units of the
 ECAL and to reject tracks which are close to the anode planes of the central jet chamber.  
Particles associated with these detector regions possess
 degraded ECAL energy and track momentum resolution respectively, resulting in a 
reduced ability to distinguish \te\ decays from other $\tau$ decays. 
 
 A total of \Vntau\ $\tau$ candidate jets pass the selection described above.  
 Estimates of non-$\tau$ backgrounds  in the $\tau$ candidate sample are summarized in table~\ref{ntau_bg}.
In cases where the data are well described by the Monte Carlo distributions the uncertainties are determined by 
the statistical error in the distributions, otherwise the uncertainties have been increased to reflect the level of 
disagreement between the data and the simulated
 distributions.  The total non-$\tau$ background was estimated to be  $\ftau=(\Vntback )\%$.
The contribution from s-channel four-fermion final states was found to be negligible for all processes 
except for $\epem \rightarrow \epem \tautau$ which contributes at the level of $\sim 0.2\%$.  However, the $\tau$ leptons 
produced in this process are added to the signal without introducing any significant measurement bias.

\renewcommand{\arraystretch}{1.1}
\begin{table}
\begin{center}
\begin{tabular}{|l|c|c|} \hline 
Source  &  $\tau$ candidate sample ($\%$) & \te\ sample ($\%$) \\  \hline 
\eeeell & \Veell         & \Veeell    \\ 
\eeqq   & \Vqqbar        &    $0$        \\
\eeee   & \Vbhabha       & \Vebhabha  \\
\eemm   & \Vmumu         &    $0$        \\ \hline
total   & \Vntback       & \Vnteback  \\ \hline
\end{tabular}
\caption[]{Estimates of non-$\tau$ backgrounds 
in the $\tau$ candidate sample and in the \te\ candidate sample. } \label{ntau_bg}
\end{center}
\end{table} 
\renewcommand{\arraystretch}{1.0}

\section{{\boldmath \te\  preselection} \label{sec:esel}}

A \te\ decay in the OPAL detector typically consists of a single track in the
 jet chamber, associated with a single ECAL cluster having essentially 
the full energy of the decay electron.   These decays are identified from the $\tau$-sample using a cut-based
 preselection followed by a likelihood selection.  The preselection  consists of a collection 
of loose selection cuts which reject $\tau$ jets which are clearly inconsistent with \te\ decays.  
  The preselection requirements are described in
the remainder of this section. 

 The kinematic variables used in this selection are based primarily on tracking and calorimetry. 
Monte Carlo modelling of the selection variables was checked over the entire momentum range using 
selected data samples of $\tau$ decays.  Additional verification of 
variable distributions for electrons was performed at high and low momenta using Bhabha and \eeeeee\ 
data samples respectively (see Section~\ref{sec:sys}).   
Several of the kinematic variables used in this analysis are ``normalized'' in order to remove the 
momentum dependence for \te\ decays.  For a variable
$K(\ptrk)$, the mean, $\mu_K (\ptrk)$, and width,
 $\sigma_K (\ptrk)$, are parameterized for \te\ decays as a function of the track 
momentum $\ptrk$.  The normalized quantity $N(K)$, defined as
\begin{equation}  N(K) \equiv \frac{K(\ptrk) -
 \mu_K (\ptrk)}{\sigma_K (\ptrk)} \mcomma \end{equation}
has a distribution which is centred at zero and with a width of unity.

\begin{figure} [tb]
\begin{center}
\mbox{\epsfig{file=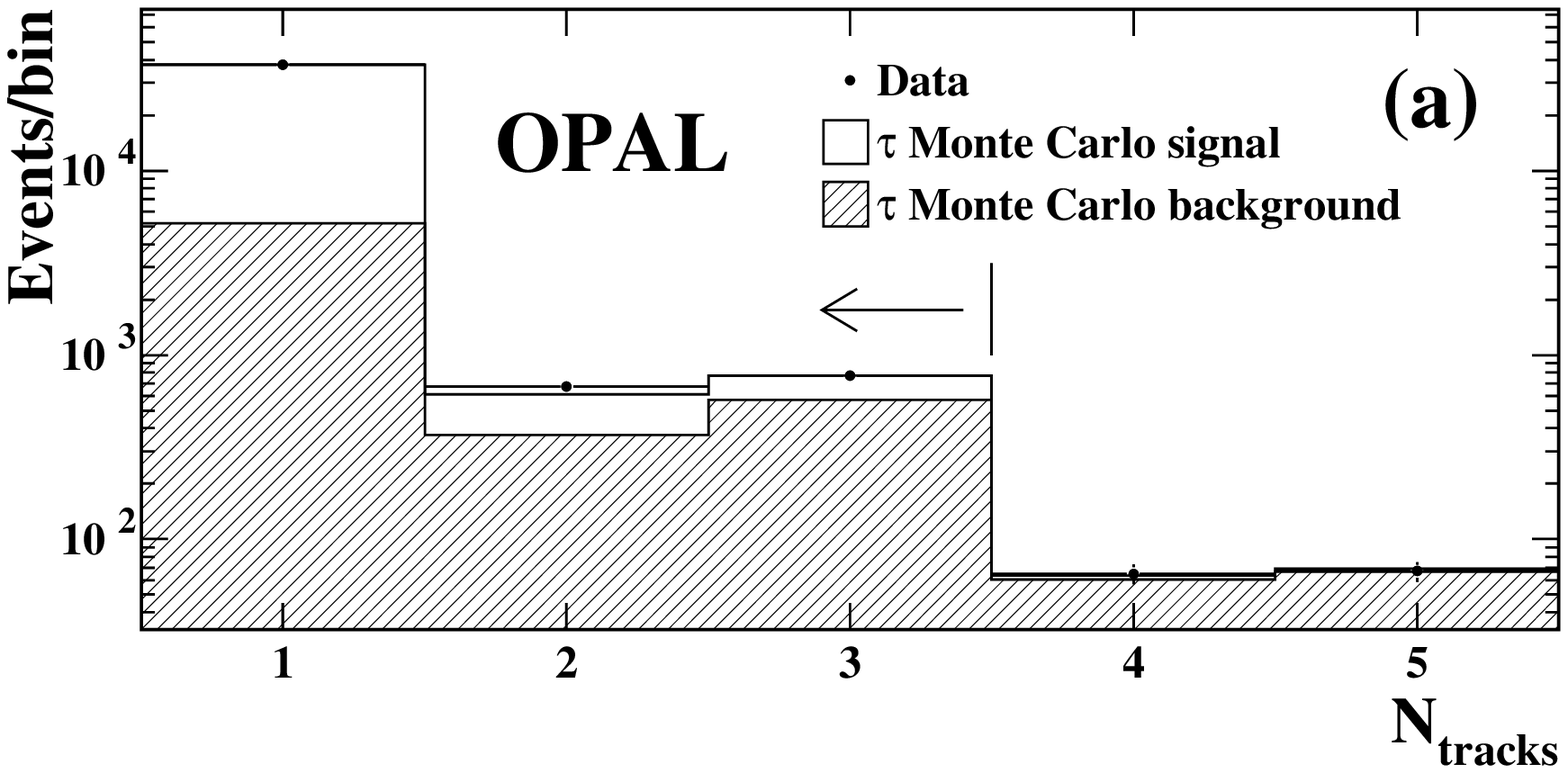,width=16.0cm,bbllx=0pt,bblly=300pt,bburx=567pt,bbury=567pt}}
\mbox{\epsfig{file=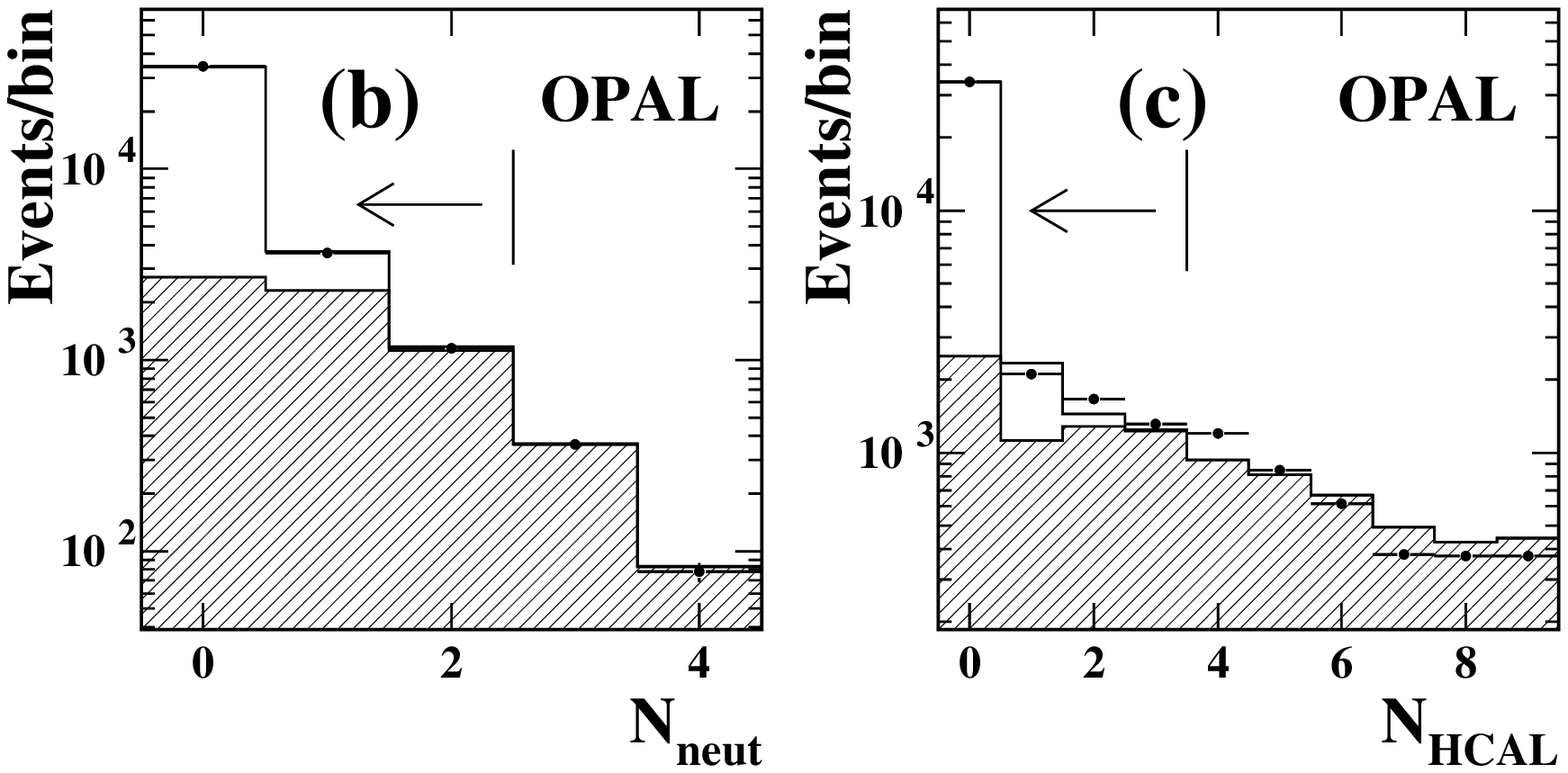,width=16.0cm,bbllx=0pt,bblly=250pt,bburx=567pt,bbury=567pt}}
\end{center}
\caption{ Comparison of data and Monte Carlo simulation of (a) the number of tracks
 $\ntrks$ in the $\tau$ jet, (b) the number of neutral clusters $\eneut$ 
and (c) the depth of penetration (in layers) into the HCAL $\hcal$,
 for $\tau$ jets which have survived all preselection cuts other than the cut on the
plotted variable.  The selected regions are indicated by the arrows.   \label{fig:ntrks} }
\end{figure}

Although most \te\ decays produce only a single track in the jet chamber, 
in approximately  $2 \%$  of these decays 
a radiated photon converts to an $\ee$ pair, producing additional tracks 
which may or may not be associated with distinct ECAL clusters. In order
 to allow for these photon conversions,  the preselection accepts jets with up to three tracks. 
If more than one track is present in a jet, the highest momentum track is assumed to be the decay
 electron and is referred to as the primary track.   
  The number of tracks in the jet ($\ntrks$) is plotted
in Figure~\ref{fig:ntrks}a for $\tau$ jets which pass all other preselection requirements.

All \te\ candidate jets which have primary track momentum $\ptrk \geq 5$~GeV are required
to have exactly one ECAL cluster associated to the track.  Jets with
$\ptrk<5$~GeV are accepted even if the primary track has no associated cluster. 
ECAL clusters which are not associated to tracks, referred to as 
neutral clusters, may be produced either by radiated photons or by photons
from \piz\ decays.  Jets containing up to two neutral clusters 
are accepted by the preselection.  The number of neutral clusters ($\eneut$) 
 is plotted in Figure~\ref{fig:ntrks}b.  

The electromagnetic shower produced by the electron from a \te\ decay is normally
 fully contained in the ECAL and little or no activity is expected in the hadron calorimeter (HCAL).
The depth of penetration of $\tau$ decay products into the HCAL is measured using
the number of sequential HCAL layers, $\hcal$, with activity associated with the $\tau$ candidate jet.  
Jets with $\hcal >3$ are rejected.   This variable is plotted in Figure~\ref{fig:ntrks}c.
The discrepancy between data and Monte Carlo in this plot is due to the modelling of
$\tau$ hadronic decays rather than \te\ decays, and its effect on the estimated background is understood.  

\begin{figure} [tb]
\begin{center}
\mbox{\epsfig{file=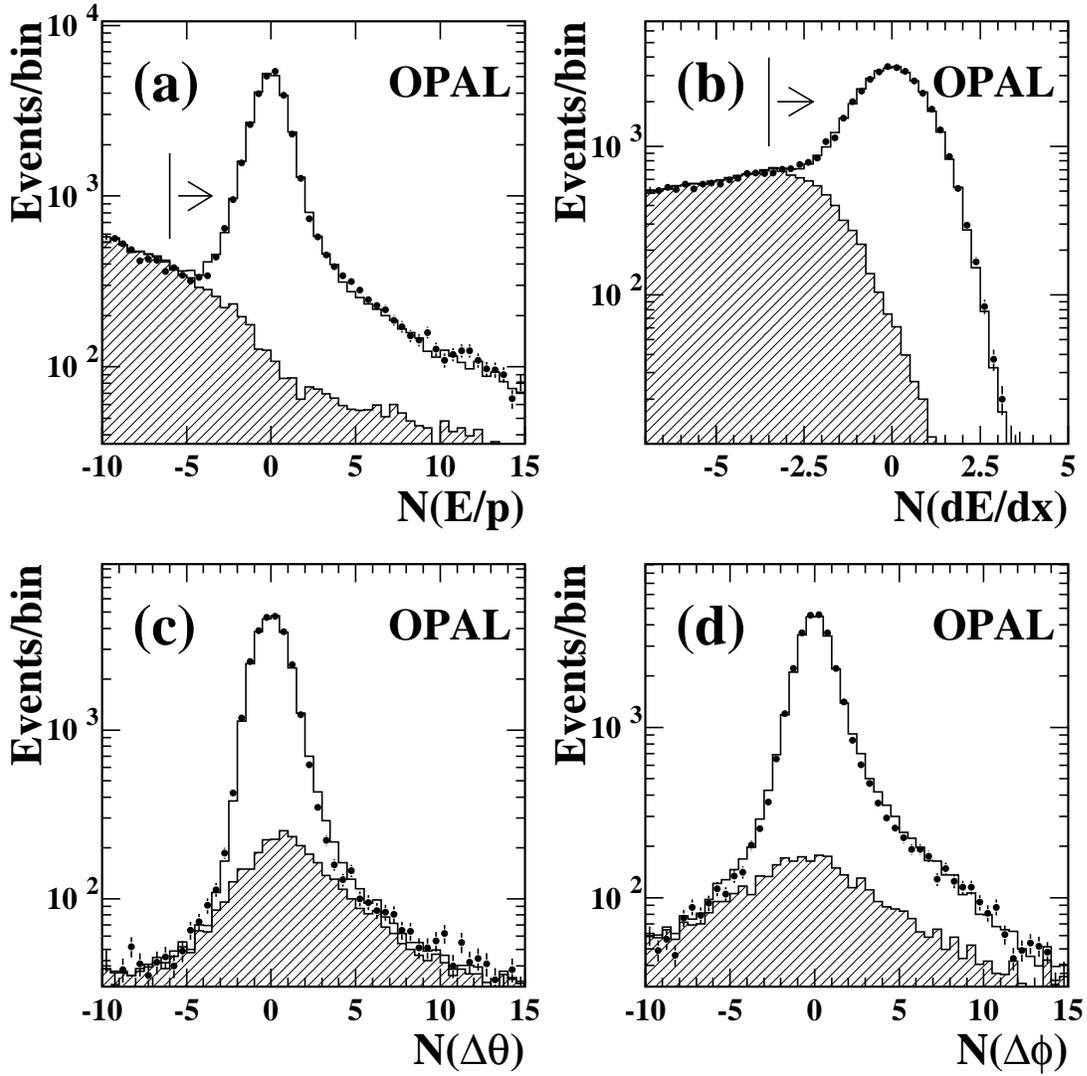,width=16.0cm,bbllx=0pt,bblly=0pt,bburx=567pt,bbury=567pt}}
\end{center}
\caption{Data and Monte Carlo simulation of the (a)  \norep\  and 
(b)  \denor\ distributions for $\tau$ jets which have survived all other 
preselection cuts.  The arrows indicate the regions accepted by the preselection
cuts.   The normalized track-cluster matching variables (c) \dtheta\ 
 and (d) \dphi\ are shown for all jets which pass the preselection.   The points represent data, while the
solid and shaded histograms are the $\tau$ Monte Carlo signal and background predictions respectively.
 \label{fig:dedx}  \label{fig:dtheta} }
\end{figure}

The ratio of the energy of the associated ECAL cluster ($\ecls$) to the momentum 
of the primary track ($\ptrk$) is typically near unity for \te\ decays, and 
less than one for other $\tau$ decays.  This ratio is normalized as described
 above to create the selection variable $\norep$ shown in
 Figure~\ref{fig:dedx}a.  A preselection cut of $\norep>-6.0$
 is applied to all $\tau$ candidate jets which have an ECAL cluster  associated with the
 primary track. 

The ionization energy deposition ($\dedx$)~\cite{dedx} of a charged particle traversing the 
central jet chamber is considered to be well measured if more than 20 
out of a possible 159 signal wires collect a measurable charge.
This criterion is satisfied for approximately $99.7\%$ of electrons produced in \te\ decays.
The normalized quantity \denor\ is constructed for all tracks with well measured $\dedx$.
Candidate jets in which the primary track possesses a  well measured $\dedx$
 and  $\denor \le -3.5 $ are rejected (see Figure~\ref{fig:dedx}b). 

All tracks other than the primary track are required to 
have properties which are consistent with electrons produced by a photon conversion.   
The quadratic sum of \denor\ of conversion-candidate tracks with well measured
$\dedx$ is required to be less than $2.5$.  The second highest momentum track in the jet
 is required to have momentum less than $4.0$~GeV.   If the jet contains three tracks, then the scalar sum
of the momenta of the second and third tracks must be less than $6.0$~GeV.

The preselection cuts described above reduce the background from hadronic $\tau$ decays to approximately $16\%$ of the sample and 
the background from  \tmu\ decays to a negligible level while maintaining a high efficiency (over $99\%$) for 
 \te\ decays.   The main \te\ selection is then accomplished by applying a likelihood selection to this sample,
further reducing the $\tau$ background to the level of $\sim 1 \%$.

\section{{\boldmath \te\  likelihood selection} \label{sec:like}}

The sample of $\tau$ decays which survive the preselection requirements described above
is used as the basis for a \te\ likelihood selection.   For this selection, the Bayesian posterior 
probability $P(e|X)$  that an event represents a \te\ decay is  estimated 
given measurements of a set $X$ of uncorrelated kinematic variables. 
The \te\ selection is accomplished by applying a cut on the estimator $P(e|X)$.
Jets which survive the preselection can be classified as either \te\ decays or \tpinpiz\ decays, so 
 likelihood functions are constructed only for these two types of decays.  The likelihood functions are based on   
reference histograms obtained from Monte Carlo simulation of a set of six selection variables.  

The first two likelihood selection variables, $\dtheta$ and $\dphi$, are normalized
 measures of the difference between the $\theta$ and $\phi$ position at which 
a charged particle enters the ECAL, as determined by tracking in the jet chamber, 
and the centroid of the associated ECAL cluster.  
Fluctuations in ECAL energy deposition by hadrons result in large variations in the
reconstructed cluster position ($\sim 9$ mrad) compared with the variations produced by electrons or
 photons (1 - 2 mrad).   This difference in position resolution is used to discriminate 
between \te\ decays and hadronic $\tau$ decays, as shown in Figure~\ref{fig:dtheta}c and~\ref{fig:dtheta}d.  

The remaining four likelihood selection variables, $\eneut$, $\hcal$, $\denor$ and $\norep$, 
were described in the previous section.   Jets which do not possess measurements of all 
likelihood selection variables are not automatically rejected.  Instead, the
likelihood is computed using the remaining variables.  This can occur, for example, if
a track with $\ptrk \le 5$~GeV does not have an associated cluster so that no measurement of
\norep\ is possible.  
The use of normalized variables implies that the selection variable reference distributions
 for \te\ decays are effectively independent of the track momentum.   The background \tpinpiz\ 
reference distributions, however, remain momentum dependent.  Reference histograms are 
therefore constructed in the three momentum bins  $\ptrk \le 5$~GeV, $5 < \ptrk \le 20 $~GeV 
and  $\ptrk >20$~GeV.  

\begin{figure} [tb]
\begin{center}
\mbox{\epsfig{file=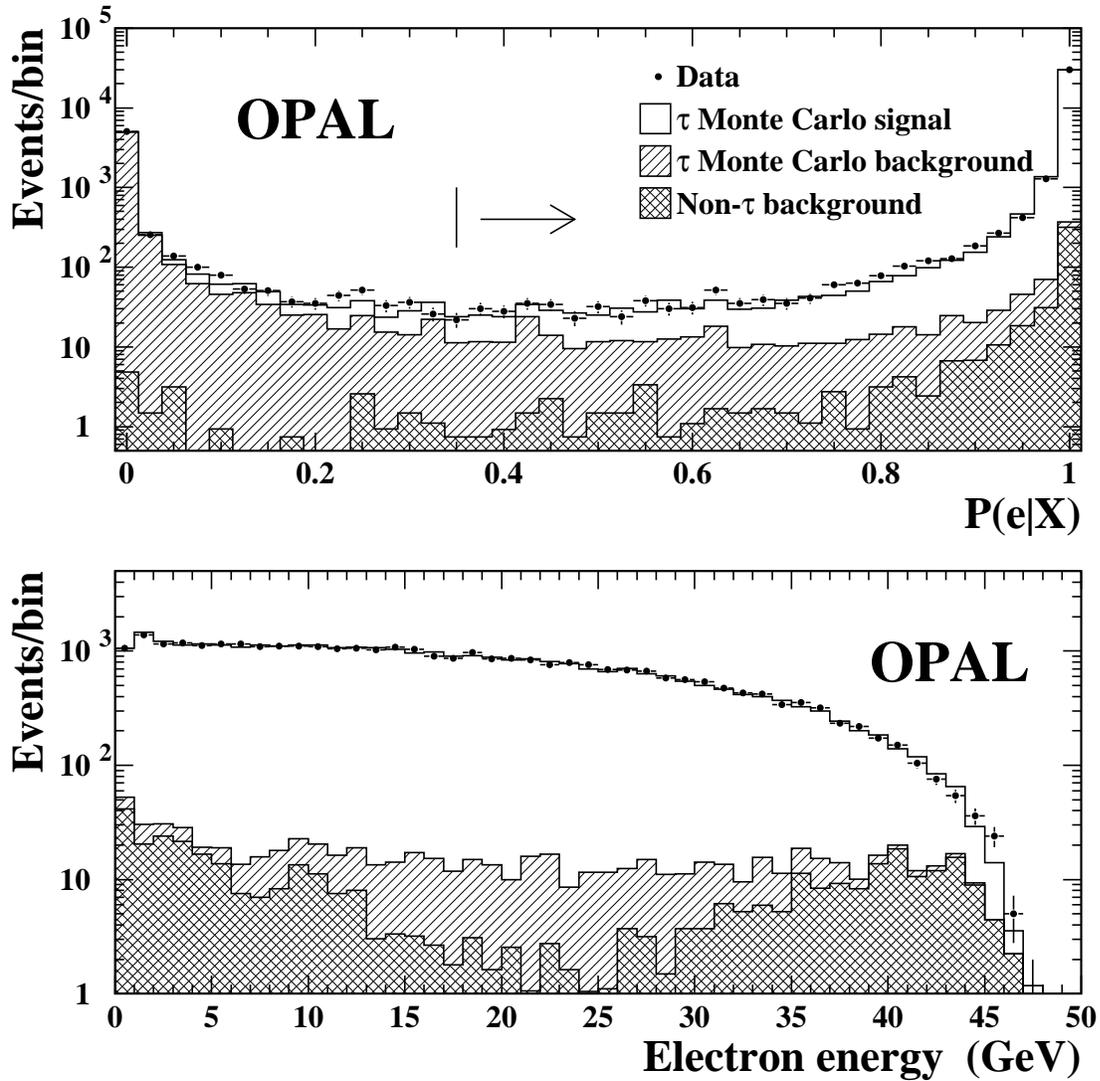,width=16.0cm,bbllx=0pt,bblly=0pt,bburx=567pt,bbury=567pt}}
\end{center}
\caption{The likelihood selection variable $P(e|X)$ is plotted for all $\tau$ jets
which pass the preselection (upper plot).  The \te\ candidate sample is composed of all jets 
in this plot with $P(e|X)>0.35$ as indicated by the arrow.   The energy distribution of
jets in the \te\ candidate sample is shown in the plot below.  Monte Carlo
predictions for the backgrounds from $\tau$ and non-$\tau$ 
sources are indicated.    \label{fig:pall} \label{fig:trk_e} }
\end{figure}

 The probability estimator $P(e|X)$ is plotted in 
Figure~\ref{fig:pall} for all $\tau$ jets which survive the preselection requirements.
The likelihood selection cut of $P(e|X)>0.35$ is optimized so as to minimize
 the total measurement uncertainty, and selects \Vnelec\ \te\ candidates 
out of the sample of $\tau$ jets.  The \te\ selection efficiency and the background due to $\tau$ other decays 
were estimated from $\tau$ Monte Carlo to be $( \Veff )\%$ and  $( \Vtback )\%$ respectively,
where the quoted uncertainties are due to the Monte Carlo statistics only.  The contributions to this
background are $(\Vbackone)\%$  from $\tpi$, $(\Vbacktwo)\%$ 
 from \tpigopiz\ and $(\Vbackthree)\%$ from other $\tau$ decays.  Systematic uncertainties
on the efficiency and background estimates are discussed in Section~\ref{sec:sys}.

Non-$\tau$ events with electrons in the
final state, such as \eeee\ and \eeeeee,  
 constitute a significant background in the \te\ candidate sample.
These non-$\tau$ background contributions to the \te\ candidate sample were evaluated 
using the same methods as the corresponding backgrounds in the $\tau$ sample.  Estimates of the contributions
 to this background are listed in Table~\ref{ntau_bg} and total $\fent = (\Vnteback )\%$ .  
The energy distribution of \te\ candidate jets is plotted in Figure~\ref{fig:trk_e}.

\section{Branching ratio results \label{sec:br} \label{sec:sys}}

The \te\ branching ratio is evaluated using the expression
\begin{equation} \brte = \left(\frac{\Necand}{\Ntcand}\right)  \cdot
 \left( \frac{1-\fet-\fent}{(1-\ftau) \cdot \eff \cdot \bias} \right) \mcomma
  \label{br_eqn} \end{equation}
where \Necand\ and \Ntcand\ are the number of selected \te\ candidates and the number
of $\tau$ candidate jets respectively, \eff\ is the efficiency for selecting \te\
 decays from the $\tau$ candidate sample and \fet\ is the background in the \te\ sample
 due to $\tau$ decays.   
 The fractional non-$\tau$ backgrounds in the $\tau$ candidate sample and in the \te\ candidate 
sample are given by \ftau\ and \fent\ respectively, and \bias\ is the $\tau$
 selection bias factor which is discussed below.  
The values of these quantities are summarized in Table~\ref{tb:br_val}.

\renewcommand{\arraystretch}{1.1}
\begin{table}
\begin{center}
\begin{tabular}{|c|c|} \hline 
Quantity  & Value \\  \hline  \hline
\Necand\  &  \Vnelec\          \\  
\Ntcand\  &  \Vntau\           \\  
\bias\    &  \Vbias\           \\ 
\ftau\    & $( \Vntback )\%$   \\ 
\fent\    & $( \Vnteback )\%$  \\ 
\eff\     & $( \Veff )\%$      \\  
\fet\     & $( \Vtback )\%$    \\ \hline 
\end{tabular}
\caption[]{Values of quantities used in the calculation of $\brte$.
Note that the quoted uncertainties on \eff\ and \fet\ are due to
 Monte Carlo statistics only.} \label{tb:br_val}
\end{center}
\end{table} 
\renewcommand{\arraystretch}{1.0}

The $\tau$ selection procedure does not in general have a uniform efficiency for selecting different
$\tau$ decay channels and therefore introduces a bias to the measured value of $\brte$.
The $\tau$ selection bias factor \bias\ measures the degree to which the $\tau$ selection
 favours or suppresses the decay \te\ relative to other $\tau$ decay channels.  It is defined
as the ratio of the fraction of \te\ decays in a sample of $\tau$ decays after the 
$\tau$ selection is applied to the fraction before the selection.  
The bias factor was evaluated using a sample of
 approximately $1.1$ million simulated $\tautau$ events.
Systematic uncertainties on \bias\ were estimated by comparing data and Monte Carlo distributions
of $\tau$ selection variables such as track and cluster 
multiplicities, and the ECAL energy scale.  Monte Carlo variables were then smeared to
cover the largest discrepancy between data and simulation and the observed shift 
in the measured bias factor was used as an estimate of the systematic uncertainty.  Additional
checks included modification of the cone finding algorithm and evaluation of the beam energy
 dependence of $\bias$.  These studies result in a bias factor estimate of $\bias = \Vbias$,
which contributes a systematic uncertainty of $ \Vbiassys \%$ to the 
branching ratio measurement. 

The non-$\tau$ backgrounds \fent\ and \ftau\ are correlated, since the background
in the \te\ candidate sample also contributes to the background in the $\tau$ sample.  
 A single systematic uncertainty of  $ \Vntbgsys \%$ is therefore assigned to the branching ratio measurement 
to account for the correlated uncertainties in the non-$\tau$ background estimates given in Table~\ref{tb:br_val}.

The systematic uncertainty due to Monte Carlo modelling
 of the likelihood selection variables was evaluated by varying
the means and widths of the four normalized variables $\denor$,
 $\norep$, $\dtheta$ and $\dphi$ around their central values and measuring the corresponding 
variation in the branching ratio result.  The ranges of variation for each variable 
 were estimated by comparing data and Monte Carlo distributions
and were chosen to cover the largest discrepancies between data 
and simulation.   The resulting shifts in the
measured branching ratio were determined independently for each varied parameter and
the uncertainty due to modelling of these variables was estimated
by adding these shifts in quadrature.  The contributions to this uncertainty from the modelling
 of the remaining two likelihood selection variables, $\eneut$ and $\hcal$, 
were estimated by varying the preselection cuts on these quantities.  
The systematic uncertainty due to Monte Carlo modelling of the six likelihood selection variables 
was estimated to be $\VMCmod \%$.   
The combined Monte Carlo statistical uncertainties on the efficiency and $\tau$-background 
estimates given in Section~\ref{sec:like} contribute an additional systematic
uncertainty of $ \VMCstat \%$ to the branching ratio measurement.  

The preselection contributes to the branching ratio systematic uncertainty predominantly
through the modelling of the preselection efficiency rather than the $\tau$ background.  
Modelling of preselection
variables was verified with data by comparing the variable distributions of selected and rejected jets
in samples in which all preselection cuts were applied other than the cut under consideration.
All selection variables are well-modelled for \te\ decays.  
 A systematic uncertainty of $ \VPresys \%$ was assigned due to the 
effect of preselection cuts which are not already covered by the selection variable modelling
 uncertainty or photon conversion uncertainties which are treated separately.  The dominant contribution 
to this uncertainty is from the associated cluster requirement for tracks with $\ptrk \geq 5$~GeV. 

The systematic uncertainty introduced by Monte Carlo modelling of
photon conversions was evaluated by modifying the preselection cuts
to reject jets possessing more than one track.
 This was found to introduce a shift of $0.005 \%$ to the measured branching ratio. 
 Jets with up to three tracks were then re-admitted to the \te\ candidate sample
 but the momentum and \denor\ cuts on the second and third tracks were not applied.  A 
significantly higher background and a slightly increased \te\ selection 
efficiency were obtained, and the reconstructed branching ratio was found to shift by  
 $0.007\%$.  A systematic uncertainty of $ \VPhosys \%$ was assigned as a result of these tests. 

The systematic uncertainty estimates from the effects discussed above
 are summarized in Table~\ref{tb:br_sys}.  Combining these values
gives a total branching ratio systematic uncertainty of $\Vpsys\%$.
As a cross check,  the likelihood selection cut was varied over the range $0.1<P(e|X)<0.8$ and the 
branching ratio result was evaluated at each value of this cut.  
The resulting variation of the branching ratio was found to be consistent with 
the estimated systematic uncertainties over the full range
of likelihood cuts.   Each of the six selection variables 
was then dropped in turn from the likelihood selection, and the posterior probability estimator 
was computed using the remaining five variables.  A cut at $0.35$ was imposed on each of 
these estimators and the six branching ratio measurements obtained in this manner
were found to be consistent within the likelihood selection systematic uncertainty 
quoted above, indicating that no individual selection variable introduces a bias 
 which exceeds the estimated uncertainty.

Modelling of selection variable distributions and
the electron identification efficiency of the \te\ selection were checked over the accessible
 momentum range using data control samples of Bhabha and two-photon \eeeeee\ events.  
 A pure data sample of electron jets was obtained from Bhabha 
events by selecting low-multiplicity events with $\etot>80$~GeV, $\acolin<0.75^\circ$
and requiring a tagged electron in the opposite jet.  The full \te\ selection was then applied to 
this data sample and the efficiency was compared with that found using the Bhabha Monte Carlo
 simulation.  A similar check was performed at low momentum using two-photon \eeeeee\ events
 selected by requiring $\acolin>20^\circ$ and a tagged electron in the opposite jet.  In both
of these data samples the overall efficiencies were found to be consistent 
with the Monte Carlo simulation at the level of $0.06\%$. 

\renewcommand{\arraystretch}{1.1}
\begin{table}
\begin{center}
\begin{tabular}{|l|c|} \hline 
Source  & Uncertainty ($\%$) \\  \hline  \hline
Bias factor               & $\Vbiassys $ \\ \hline  
Likelihood selection modelling & $ \VMCmod $ \\ \hline 
Non-$\tau$ backgrounds    & $ \Vntbgsys $ \\ \hline 
Preselection              & $ \VPresys $ \\ \hline
Monte Carlo statistics    & $ \VMCstat $ \\ \hline
Photon conversions        & $ \VPhosys $ \\ \hline  \hline 
Total systematic uncertainty  & $ \Vpsys $ \\ \hline   
\end{tabular}
\caption[]{Contributions to the total branching ratio absolute systematic uncertainty.} \label{tb:br_sys}
\end{center}
\end{table}
\renewcommand{\arraystretch}{1.0}

 Substituting the values from Table~\ref{tb:br_val} into equation~\ref{br_eqn} and using the 
systematic uncertainty estimate above gives a branching ratio measurement of
\beq \brte=\left( \Vbr \pm  \Vstat \mbox{ (stat)} \pm \Vsys \mbox{ (syst)} \right)\%  \mperiod \eeq 
This result is the most precise measurement of \brte\ to date, and is in good agreement with other
 measurements of this quantity~\cite{pdg}.

\section{Lepton universality tests \label{sec:univ}}

In the Standard Model, the leptonic partial decay widths of the $\tau$, including electroweak radiative corrections 
of order $\alpha$ and neglecting neutrino masses, are given by~\cite{tsai,marciano}
\begin{equation}  \Gamma(\tl) = \frac{g_\tau^2 g_l^2}{(8m_{\mathrm{W}}^2)^2} \frac{m_\tau^5}{96\pi^3}
f \left( \frac{m_l^2}{m_\tau^2} \right)(1+\delta_{RC}^\tau)  \mcomma \label{eq:width}
  \end{equation}
where $m_l$ is the mass of the lepton ($l=\rm{e},\mu $), $m_\tau$ is the 
mass of the $\tau$ lepton, $m_{\rm{W}}$ is the mass of the $\Wpm$ boson
 and $f(x) = 1 - 8x +8x^3 -x^4 -12x^2 \ln x$ is a phase-space correction factor.  The term
\begin{equation} (1+\delta_{RC}^\tau) =  \left[ 1+ \frac{3m_\tau^2}{5m_{\mathrm{W}}^2} \right] 
\left[ 1+ \frac{\alpha(m_\tau^2)}{2\pi} \left( \frac{25}{4} -\pi^2 \right) \right] 
\label{eq:RC}\end{equation}
represents leading order $\Wpm$ propagator and radiative corrections, where
 $\alpha(m_\tau^2) =1/133.29$ is the fine structure constant at the $\tau$
 mass scale.   The coupling constants $g_\tau$ and $g_l$ 
describe the strength of the coupling of the $\Wpm$ to the different lepton generations. 
Lepton universality requires that $g_{\rm e}$, $g_\mu$ and $g_\tau$ are identical.

The ratio of $g_\mu/g_{\rm e}$ can be evaluated by comparing the $\tau$ decay widths 
to electrons and muons using equation~\ref{eq:width}, giving
\begin{equation} \left(\frac{g_\mu}{g_{\rm e}} \right)^2 = 
\frac{\brtmu}{\brte} \frac{f(m_{\mathrm{e}}^2/m_\tau^2)}{f(m_\mu^2/m_\tau^2)} \mperiod
\end{equation}
The phase-space correction factors have the values $f(m_{\mathrm{e}}^2/m_\tau^2) = 1.0000$ and 
$f(m_\mu^2/m_\tau^2) = 0.9726$.
Using the OPAL \tmu\ branching ratio measurement $\VObrtmu$~\cite{clayton}, 
one obtains $g_\mu/g_{\rm e}= \Vgmuge$, in
 good agreement with the hypothesis of e - $\mu$ universality.

\begin{figure} [tb]
\begin{center}
\mbox{\epsfig{file=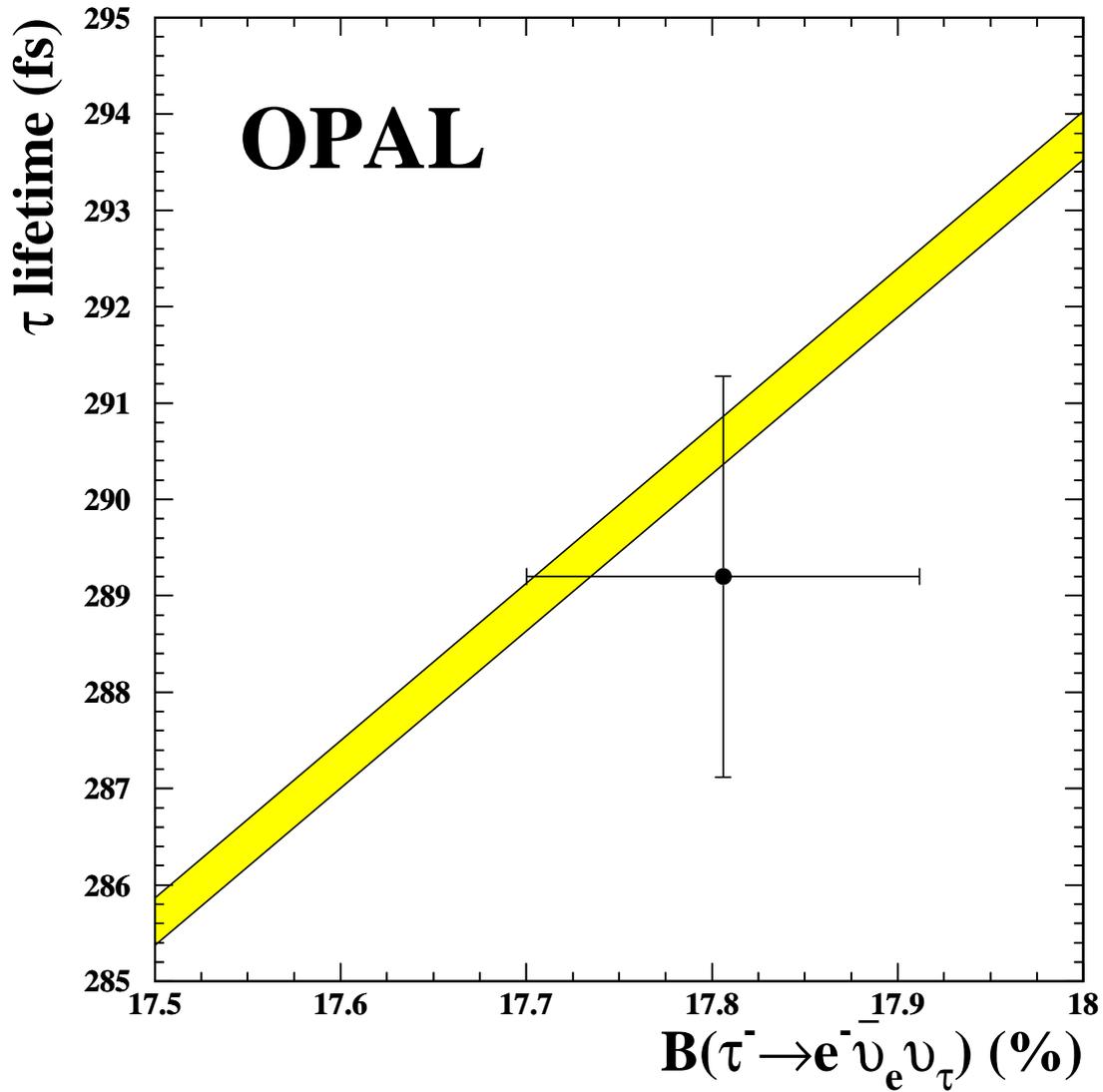,width=16.0cm,bbllx=0pt,bblly=0pt,bburx=567pt,bbury=567pt}}
\end{center}
\caption{The OPAL $\tau$ lifetime is plotted against the \te\ branching ratio.  
The shaded band is the Standard Model prediction under the assumption of lepton universality,
 and its width represents the uncertainty due to the measured $\tau$ mass.      \label{fig:br_vs_ttau}}
\end{figure}

 A test of $\mu$ - $\tau$  universality can be made by  
comparing the partial widths for \te\ and \mue\ decays.
The \mue\ width is given by an expression equivalent to equation~\ref{eq:width}.  
The ratio of the couplings $g_\tau$ and $g_\mu$ is then given by 
\begin{equation}  
\frac{\Gamma(\te)}{\Gamma(\mue)} = \frac{g_\tau^2 \; m_\tau^5 \;
f\left(m_{\mathrm{e}}^2 / m_\tau^2 \right) (1+\delta_{RC}^\tau)}{ g_\mu^2  \; m_\mu^5  \;
f\left(m_{\mathrm{e}}^2 / m_\mu^2 \right)(1+\delta_{RC}^\mu)} \mcomma
\end{equation} 
where $(1+\delta_{RC}^\mu)$ represents the $\Wpm$ propagator and radiative corrections to 
the $\mue$ width analogous to equation~\ref{eq:RC}.  
Substituting numerical values for the correction factors gives
\begin{equation} \left(\frac{g_\tau}{g_\mu}\right)^2 =  0.9996~ \frac{\tau_\mu}{\tau_\tau} 
 \frac{m_\mu^5}{m_\tau^5}~\brte \mperiod
\end{equation}
Using the OPAL value for the $\tau$ lifetime, $\tau_{\tau} = \VOtlife $~\cite{opal_tlife}, 
the BES Collaboration value for the $\tau$ mass, $\VMtau$~\cite{bes}, and
Particle Data Group~\cite{pdg} values for the muon mass and lifetime, $\tau_{\mu}$, gives the
result $g_\tau/g_\mu = \Vgtaugmu$.  The OPAL $\tau$ lifetime and
\brte\  are compared with the Standard Model prediction in Figure~\ref{fig:br_vs_ttau}.

\section{\boldmath Measurement of \as\ \label{sec:alphas}}

A measurement of \as\  can be extracted from the ratio $R_\tau$ of the hadronic decay width
to the electronic decay width:
\begin{equation}  R_\tau \equiv \frac
{\Gamma(\mbox{$\tau^{-}\rightarrow \mbox{ hadrons } \nu_\tau$})}
{\Gamma(\te)} \mperiod
 \label{eq:rtau} \end{equation}
Significant deviations from the parton level prediction $R_\tau \simeq 3$ can be
 accounted for in terms of QCD dynamics~\cite{lamyan,braaten_2,bnp,diberder} and are sufficiently 
large to allow a measurement of  $\as$. 
Perturbative and non-perturbative corrections to the parton level prediction
can be organized in powers of $1/m_\tau^2$ using the short-distance Operator Product
 Expansion~\cite{SVZ}, giving 
\begin{equation} R_\tau = 3 \left(|\vud|^2 + |\vus|^2 \right) S_{EW}
\left\{ 1 + \delta_{EW} + \delta^{(0)} + \sum_{D=2,4,...} \delta^{(D)} 
\right\} \label{eq:rtau_exp}  \mcomma \end{equation}
where $\delta^{(D)}$ are QCD condensates which are suppressed by $1/m_\tau^{D}$.  The factors
$S_{EW} = \Vsew$~\cite{marciano} and $\delta_{EW} = \Vdew$~\cite{li} are electroweak 
corrections, and $\vud= \Vvud$ and $\vus=\Vvus$ are CKM matrix elements~\cite{pdg}. 
Perturbative QCD corrections contribute at dimension $D=0$ and have been 
calculated to be~\cite{braaten_2}
\begin{equation} \delta^{(0)} = \left( \frac{\asmt}{\pi} \right) 
+5.2023 \left( \frac{ \asmt }{\pi}\right)^2 
+26.366\left( \frac{ \asmt}{\pi}\right)^3
+ {\mathcal{O}} (\as ^4) \mperiod \label{eq:pert}\end{equation}
The coefficient of the ${\mathcal{O}}(\as ^4)$ term is $(78.00 + K_4)$, where the $K_4$
coefficient has been estimated to be $K_4 \approx 25 \pm 50$~\cite{diberder,pich_9703}.
The theoretical uncertainty in the perturbative expansion is taken to be due only to the $K_4$
uncertainty.  Quark mass corrections contribute at dimension $D=2$ and total
 $\delta^{(2)} = \VDtwocor$. 
Non-perturbative QCD corrections enter at dimension $D \geq 4$ and have been estimated to total
$\Vnpert$~\cite{bnp,tau94_narison}.  Recent experimental measurements of
the moments of the $\tau$ hadronic spectral functions have confirmed the size of these
condensates~\cite{aleph_spec,sven}.  

Under the assumption of lepton universality, $R_\tau$ can be expressed 
in terms of the \te\ branching ratio only:
\begin{equation}  R_\tau = \frac{1- \brte \cdot (1.9726)}{\brte} \mcomma
\label{eq:rtau_be}  \end{equation}
where the numerical factor is from the phase space correction factor in Section~\ref{sec:univ}.
Substituting the branching ratio result from Section~\ref{sec:br} 
gives the experimental value $R_\tau = \VRtau \pm \VRterr $. 
A value of \asmt\ can be extracted from this result using equations~\ref{eq:rtau_exp}
 and~\ref{eq:pert}, yielding
\begin{equation} \asmt = \Vastau \pm \Vasteexp \mbox{ (exp)} 
\pm \Vasteth \mbox{ (theory)\hspace{0.4cm}.}  
\label{eq:as_mtau}\end{equation}
Contributions to the theoretical uncertainty, listed in Table~\ref{tb:as_errors},
are dominated by the renormalization-scheme and scale dependences.  
The scale dependence is taken from~\cite{sven}, while the
 renormalization scheme dependence was estimated
 in~\cite{raczka} by evaluating the effect of changing from the
 $\overline{\mathrm{MS}}$ scheme to a scheme using the principle of minimal sensitivity (PMS).   

The value of \asmt\ is  evolved to the \zo\ mass scale for comparison with other measurements.
The $\beta$ coefficients of the renormalization group equations 
have recently been calculated up to four loops in the $\overline{\mathrm{MS}}$ scheme~\cite{beta3}.
Following the procedure described in~\cite{run_as}, running is performed at four loops, with 
 three-loop matching conditions used to pass the heavy quark mass thresholds.  
 This procedure has been shown to introduce an 
uncertainty of \Vzrun\ to the extracted value of \asmz~\cite{run_as}.  
Evolving the result from expression~\ref{eq:as_mtau} gives 
\beq \asmz = \Vasz \pm \Vaszeexp \mbox{ (exp)} \pm \Vaszeth \mbox{ (theory)  ,} \eeq  
 which is in good agreement with determinations of \asmz\ using measurements at other
energy scales~\cite{pdg}.

\renewcommand{\arraystretch}{1.20}
\begin{table}
\begin{center}
\begin{tabular}{|l|c|c|c|} \hline 
Source  & Uncertainty & $\Delta \asmt$ & $\Delta \asmz$ \\ \hline \hline
Scale dependence    & $ \mu^2/m_\tau^2 \rightarrow 0.4 - 2.0$  & $ \Vtsd $  & $ \Vzsd $  \\ \hline 
RS dependence    & $\overline{\mathrm{MS}} \rightarrow \mathrm{PMS}$ & $ \Vtrs $ & $ \Vzrs $  \\  \hline  
$K_4$ dependence    & $K_4 = 25 \pm 50$         & $ \Vtkf $ & $ \Vzkf $  \\  \hline 
$\delta^{(D)}$ corrections ($D\geq 2$)     & \Vnpertcor\    & $ \Vtnp $  & $ \Vznp $  \\  \hline 
Electroweak corrections & $1.0194 \pm 0.0040$       & $ \Vtew $ & $ \Vzew $ \\  \hline 
Running $\asmt \rightarrow \asmz $ & - & - & $ \Vzrun $ \\  \hline  \hline 
Total theory uncertainty   & - & $ \Vasteth $ &  $ \Vaszeth $ \\  \hline 
\end{tabular}
\caption{Theoretical uncertainties on \asmt\ and $\asmz$. \label{tb:as_errors} }
\end{center}
\end{table}
\renewcommand{\arraystretch}{1.00}

\section{Conclusions \label{sec:conc}}

Data collected by the OPAL experiment at energies 
near the \zo\ resonance have been used to determine the \te\ branching ratio.
A total of \Vnelec\ candidate \te\ decays were identified from a sample of
 \Vntau\ $\tau$ decays, using a likelihood-based selection, to give a branching ratio
 measurement of 
$\brte = (\Vbr \pm \Vstat \mbox{ (stat)} \pm \Vsys \mbox{ (syst) })\% $.
This result supersedes previous OPAL \te\ branching ratio
 measurements~\cite{randy,clayton} and is the most precise measurement of this quantity to date.  
This result has been combined with measurements of other quantities
to test the universality of e - $\mu$ and $\mu$ - $\tau$ charged-current
 couplings to a precision of better than $1\%$.
In addition, the strong coupling constant \asmt\
has been extracted from $\brte$ yielding 
$\asmt = \Vastau \pm \Vasteexp \mbox{ (exp)} \pm \Vasteth \mbox{ (theory)}$. 
Evolving this value to the \zo\ mass scale gives 
$\asmz = \Vasz \pm \Vaszeexp \mbox{ (exp)} \pm \Vaszeth \mbox{ (theory)}$.  

\section*{Acknowledgements}

\noindent We particularly wish to thank the SL Division for the efficient operation
of the LEP accelerator at all energies
 and for their continuing close cooperation with
our experimental group.  We thank our colleagues from CEA, DAPNIA/SPP,
CE-Saclay for their efforts over the years on the time-of-flight and trigger
systems which we continue to use.  In addition to the support staff at our own
institutions we are pleased to acknowledge the  \\
Department of Energy, USA, \\
National Science Foundation, USA, \\
Particle Physics and Astronomy Research Council, UK, \\
Natural Sciences and Engineering Research Council, Canada, \\
Israel Science Foundation, administered by the Israel
Academy of Science and Humanities, \\
Minerva Gesellschaft, \\
Benoziyo Center for High Energy Physics,\\
Japanese Ministry of Education, Science and Culture (the
Monbusho) and a grant under the Monbusho International
Science Research Program,\\
Japanese Society for the Promotion of Science (JSPS),\\
German Israeli Bi-national Science Foundation (GIF), \\
Bundesministerium f\"ur Bildung, Wissenschaft,
Forschung und Technologie, Germany, \\
National Research Council of Canada, \\
Research Corporation, USA,\\
Hungarian Foundation for Scientific Research, OTKA T-016660, 
T023793 and OTKA F-023259.\\


%
\newpage

\end{document}